\documentclass{aa}
\usepackage{newtxtext,newtxmath}
\usepackage[T1]{fontenc}
\usepackage{ae,aecompl}
\usepackage{xcolor}
\bibpunct{(}{)}{;}{a}{}{,}
\usepackage[colorlinks=true, citecolor=blue]{hyperref}
\usepackage{multirow}
\makeatletter
\renewcommand*\aa@pageof{, page \thepage{} of \pageref*{LastPage}}
\makeatother
\usepackage{etoolbox}
\usepackage{graphicx}	
\usepackage{amsmath}	
\usepackage{verbatim}
\usepackage{units}
\usepackage{pdflscape}
\usepackage{svg}
\usepackage{bm}
\usepackage{stfloats}
\usepackage{float}
\usepackage{multicol}
\usepackage{hyperref}
\usepackage{multirow}
\usepackage{wasysym}
\usepackage{subcaption}
\usepackage{dcolumn}
\newcolumntype{d}[1]{D{.}{.}{#1}}
\newcommand\mc[1]{\multicolumn{1}{c}{#1}}

\newcommand{\mdotsun}{$\dot{\mathrm{M}}_{\odot}$}
\newcommand{\mdot}{$\dot{\mathrm{M}}$}

\begin{document}

\title{Catching the wisps: Stellar mass-loss limits from low-frequency radio observations} 

\author{S.~Bloot$^{1,2}$\thanks{Corresponding author \email{bloot@astron.nl}}, H.~K.~Vedantham$^{1,2}$, R.~D.~Kavanagh$^{1,3}$, J.~R.~Callingham$^{1,3}$, B.~J.~S.~Pope$^{4,5,6}$}
\authorrunning{S.~Bloot et al.}
\institute{$^{1}$ASTRON, Netherlands Institute for Radio Astronomy, Oude Hoogeveensedijk 4, Dwingeloo, 7991\,PD, The Netherlands\\
$^{2}$Kapteyn Astronomical Institute, University of Groningen, P.O. Box 800, 9700 AV, Groningen, The Netherlands\\
$^{3}$Anton Pannekoek Institute for Astronomy, University of Amsterdam, 1098 XH, Amsterdam, the Netherlands\\
$^{4}$School of Mathematics and Physics, The University of Queensland, St Lucia, QLD 4072, Australia\\
$^{5}$Centre for Astrophysics, University of Southern Queensland, West Street, Toowoomba, QLD 4350, Australia\\
$^{6}$School of Mathematical \& Physical Sciences, 12 Wally’s Walk, Macquarie University, NSW 2113, Australia\\
}
\date{Received XXX; accepted YYY}

\label{firstpage}

\abstract
{The winds of low-mass stars carry away angular momentum and impact the atmospheres of surrounding planets. Determining the properties of these winds is necessary to understand the mass-loss history of the star and the evolution of exoplanetary atmospheres. Due to their tenuous nature, the winds of low-mass main-sequence stars are difficult to detect. The few existing techniques for measuring these winds are indirect, with the most common inference method for winds of low-mass stars being astrospheric Lyman-$\alpha$ absorption combined with complex hydrodynamical modelling of the interaction between the stellar wind and the interstellar medium. Here, we employ a more direct method to place upper limits on the mass-loss rates of low-mass stars by combining observations of low-frequency coherent radio emission, the lack of free-free absorption, and a simple stellar wind model. We determine upper limits on the mass-loss rate for a sample of 19 M\,dwarf stars detected with the LOFAR telescope at 120--168\,MHz, reaching a sensitivity within an order of magnitude of the solar mass-loss rate for cold stars with a surface magnetic field strength of $\sim$100\,G. The sensitivity of our method does not depend on distance or spectral type, allowing us to find mass-loss rate constraints for stars up to spectral type M6 and out to a distance of 50\,pc, later and farther than previous measurements. With upcoming low-frequency surveys with both LOFAR and the Square Kilometre Array, the number of stars with mass-loss rate upper limits determined with this method could reach $\sim$1000.
}
\maketitle
\section{Introduction}
\noindent Stellar winds carry angular momentum away from stars, leading to a decrease in their rotation rate and dynamo-induced magnetic activity \citep[e.g.][]{2012AJ....143...93R, 2012LRSP....9....1R,2014MNRAS.441.2361V}. The stellar wind also heavily impacts orbiting planets, eroding their atmospheres and changing their composition (see e.g. \citet{owen2019} for a review). The majority of conventional habitable-zone planets orbit M\,dwarfs \citep{2013ApJ...767...95D,2013ApJ...767L...8K}, where the habitable zone is closer to the star than for the Sun. This implies that the stellar wind could have a more significant impact on habitability than for the Earth-Sun system, which is a major uncertainty for estimates of the number of habitable planets in the universe \citep[e.g.][]{Shields2016,Kipping2021}.

Stellar winds can be characterised by their mass-loss rate, which determines the density and momentum of the wind over time. The majority of stellar mass-loss rates for low-mass stars have been inferred using astrospheric Lyman-$\alpha$ absorption \citep[e.g.][]{wood2002}. This method requires knowledge of the density and neutral fraction of the interstellar medium (ISM), limiting its applicability to stars within $\sim$7\,pc \citep{wood2004}.
Stellar mass-loss rates can also be determined with X-ray observations of the astrosphere, using soft X-ray emission resulting from the charge
exchange between heavy ions in the stellar wind and cold neutrals in the
ISM \citep{2024arXiv240414980K}. With these X-ray measurements, the wind density can be determined directly, albeit only for nearby stars (within 5-10\,pc) due to the faintness of the signal and the high spatial resolution required. 
Three stars have been successfully studied with both Ly$\alpha$ and X-ray measurements, where the two techniques can disagree by more than an order of magnitude \citep{2024arXiv240414980K, wood2002,Wood_2005, Wood_2014}.

On young, rapidly rotating stars, \citet{2019MNRAS.482.2853J} have estimated mass-loss rates using slingshot prominences, where material flows from the stellar surfaces into closed magnetic loops. The density at the base of these loops is independent of the prominence structure, and can therefore be used to determine the mass-loss rate. The sample of stars available for this method is inherently limited, as it requires a detection of prominences on the star.

Another method can be used to infer the stellar wind of a low-mass star if it has a companion white dwarf. In that case, the stellar wind from a low-mass star accretes onto the white dwarf, resulting in detectable metals in the white dwarf spectrum that can be used to determine the mass-loss rate of the low-mass star \citep{white_dwarf_wind}. However, the exact accreting efficiency of white dwarfs is still unknown, implying such a method can only provide lower limits on the mass-loss rate.

At radio frequencies of $\sim$10-100\,GHz, the mass-loss rate can be measured by detecting free-free emission from the wind. The flux density of free-free emission scales with the density and the temperature of the wind, and therefore traces the mass-loss rate. However, stars often also produce chromospheric thermal emission and coronal synchrotron emission at these frequencies \citep[e.g.][]{dulk1985}, which are difficult to disentangle from free-free emission. Previous work in this area has produced upper limits \citep[e.g.][]{fichtinger}, where the emission mechanism cannot be uniquely determined or no radio emission was detected from the star.

Here we constrain the mass-loss rate by the observed absence of free-free absorption (FFA) in the coronae of radio-detected stars. The FFA optical depth is proportional to the integral of the density squared along the line of sight to the emitter \citep[e.g.][]{Rybicki_Lightman}. If the radio emission from the star is detected and therefore not catastrophically absorbed, it provides a direct upper limit on this integral. The constraint on the integral of the density can be converted to an upper limit on the mass-loss rate with a simple model for the radial density profile if the height of the emitter from the stellar surface can be determined.

While the FFA method has been employed before \citep{1996ApJ...462L..91L}, it has been underused because the height of the emitter from the stellar surface was unknown. Furthermore, most previous radio detections of stars were at relatively high frequencies of $\sim$1-10\,GHz \citep[e.g.][]{villadsen2019, driessen2024}, while the constraint from FFA is much stronger at lower frequencies. With the renaissance of low-frequency radio astronomy allowing for more detections of radio emission from stars with confirmed emission mechanisms \citep[e.g.][]{davis-2019,2021A&A...653A.101F, toet,crdra,callingham2021, vlotss, yiu2024}, combined with the availability of Zeeman-Doppler Imaging \citep[ZDI; e.g.][]{1997A&A...326.1135D}, we can now use this method to find competitive mass-loss rate constraints for stars for which other methods would be inadequate.

\section{Mass-loss limits via free-free absorption}
\noindent To determine the mass-loss rate of a star based on the lack of strong free-free absorption, we first need a detection of radio emission. The detection implies that the radiation was not catastrophically free-free absorbed, regardless of other propagation effects. Here, we focus on emission produced through the electron-cyclotron maser instability (ECMI).
ECMI is generated at the fundamental or second harmonic of the cyclotron frequency. The cyclotron frequency, $\nu_{\mathrm{B}}$, depends on the ambient magnetic field strength, $B$ according to
\begin{equation}
    \label{eq:cyc_freq}
    \nu_{\mathrm{B}} \approx 2.8 \times 10^6 \left(\frac{B}{\rm Gauss}\right)\,\,{\rm Hz}.
\end{equation}
If the plasma frequency at the source location is too high, ECMI will not be produced. However, we do not remove scenarios where this occurs in our modelling, to account for the possibility that the emission is produced in small-scale cavities in the wind.

We approximate the radial profile along the magnetic axis of the large-scale field of the star as a dipole:
\begin{equation}
    B(R)=B_0\left(\frac{R}{R_*}\right)^{-3},
\end{equation}
where $B_0$ is the surface magnetic field strength, $R_*$ is the stellar radius, and $R$ is the distance from the centre of the star. Beyond the radius where the magnetic pressure is equal to the dynamic pressure of the wind, we assume the field lines open up, and that the field evolves as $B(R)\propto\left(\frac{R}{R_*}\right)^{-2}$. We also assume that the emission is produced at the second harmonic ($\nu=2\nu_{\mathrm{B}}$). This is a conservative assumption, as it places the emitter as far from the stellar surface as possible such that it suffers the least possible free-free absorption. With this model for the magnetic field, we can map the emission frequency to a height from the stellar surface where the emission is produced. 
If the measured large-scale field of the star is too weak to produce emission at the observed frequency, we assume the emission must be produced in small-scale magnetic structures on the surface. 

\begin{figure}
    \centering
    \includegraphics[trim={3cm 4cm 3cm 4cm},clip,width=0.99\columnwidth]{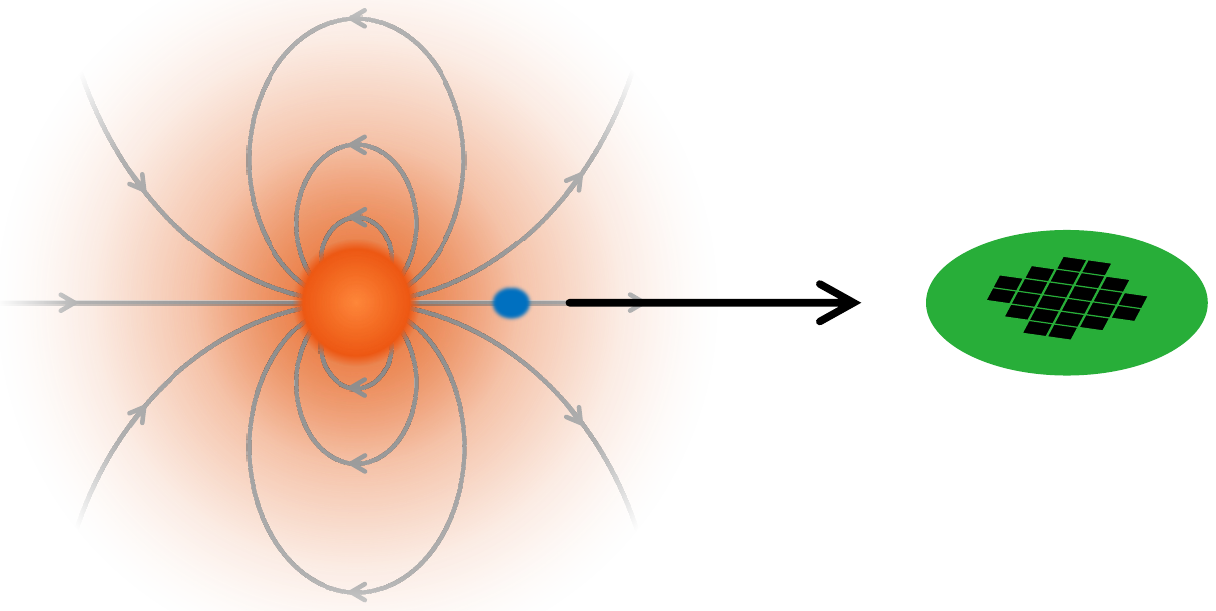}
    \caption{A diagram illustrating the geometry we assume in the free-free absorption modelling. The magnetic field dipole is shown in grey. We place the emitter above the magnetic pole and assume the emission propagates radially outwards along the line of sight, to ensure the most conservative mass-loss upper limit of all possible geometries. The emission region (shown in blue) is assumed to be as close to the observer as possible, to minimise the free-free absorption along its path (black arrow) to the detector (shown here as a representation of LOFAR).}
    \label{fig:diagram}
\end{figure}

ECMI originating in the large-scale field is expected to be produced around the magnetic poles, whereas emission from small-scale magnetic loops could be produced at any latitude where active regions are present. Regardless, ECMI emission is expected to be beamed along the surface of a cone whose axis is parallel to the ambient magnetic field \citep[e.g.][]{dulk1985}. The cone opening angle and the magnetic latitude of the emitter are not known a~priori. We therefore make the most conservative assumption--- that the emission travels radially outward, which is the path of least free-free absorption through the corona in a spherically symmetric wind. This geometry is illustrated in Fig\,\ref{fig:diagram}. The reason we adopt these assumptions, even if they are not necessarily accurate, is that any other ray path will result in more stringent mass-loss constraints.

We determine the wind structure using an isothermal Parker wind model \citep{parker}, based on the mass, radius, and coronal temperature of the star. We solve the momentum equation for the Parker wind model, given by \begin{equation}
\frac{1}{v}\frac{dv}{dR}=\left(\frac{2 v_s^2}{R} -\frac{G M_*}{R^2} \right)/(v^2-v_s^2)
\end{equation} where $v$ is the velocity, $v_s$ is the sound speed, $R$ is the radial distance, $G$ is the gravitational constant, and $M_*$ is the stellar mass. We select the solution that corresponds to an outflowing wind.
The sound speed is given by \begin{equation}
    v_{s}=\sqrt{\frac{k_B T}{\mu m_p}},
\end{equation}
where $k_B$ is the Stefan-Boltzmann constant, $T$ is the temperature, $\mu$ is the mean molecular weight (taken to be 0.6), and $m_p$ is the mass of a proton. The critical radius at which this speed is reached is  \begin{equation}
    R_c=\frac{GM_*}{2 v_s^2}.
\end{equation} We combine the resulting velocity profile with a mass-loss rate to model the radial density profile according to \begin{equation}
    \rho(R)=\frac{\dot{{M}}}{v(R) 4 \pi R^2},
\end{equation}
where $\dot{{M}}$ is the mass-loss rate and $v(R)$ is the velocity profile of the wind. 

We compared the results from the Parker model to the Weber-Davis model \citep{1967ApJ...148..217W} and found that the two models produce similar results for these stars. This is likely because the radial density and magnetic field structure in the two models diverge at large radii, whereas the bulk of the absorption happens at much smaller radii.

We compute a density structure for a grid of potential total base densities, ranging from $10^6$\,cm$^{-3}$ to $10^{18}$\,cm$^{-3}$. For each base density, we calculate the optical depth from the emission region out to 100 solar radii, at which point the free-free absorption is negligible at our frequencies of interest. We calculate the optical depth $\tau$ using 
\begin{equation}
    \tau = \int 0.018 T^{-1.5} n_e n_i \nu^{-2} g_{\mathrm{ff}} \frac{1}{\sqrt{1-\nu_p^2/\nu^2}}\mathrm{d}l
\end{equation}
where $T$ is the temperature in K, $n_e$ is the electron number density in cm$^{-3}$, $n_i$ is the ion number density in cm$^{-3}$, $\nu$ is the frequency at which the emission is observed, $\nu_p$ is the plasma frequency, $l$ is the distance along the line of sight, and $g_{\mathrm{ff}}$ is the Gaunt factor \citep{Rybicki_Lightman}. We calculate the Gaunt factor from tables given by \citet{2014MNRAS.444..420V}. The factor in the denominator is the refractive index of plasma, which takes into account the fact that close to the plasma frequency the group velocity deviates significantly from the speed of light \citep{1996ASSL..204.....Z}. We use the refractive index of isotropic plasma as an approximation. We determine the isothermal wind temperature from the coronal temperature from the empirical relation found by \citet{johnstone} for main-sequence low-mass stars, based on the correlation between the surface X-ray flux of the star and its coronal temperature, given by \begin{equation}
    T_{\mathrm{cor}} \approx 0.11F_X^{0.26},
\end{equation}
where $T_{\mathrm{cor}}$ is the coronal temperature in MK and $F_X$ is the surface X-ray flux in erg\,s$^{-1}$\,cm$^{-2}$. We estimate the base temperature of the wind by scaling the coronal temperature down by a factor of 1.36, determined by comparing the solar coronal and wind temperature, following \citet{2018MNRAS.476.2465O}.
For LP\,169-22, we only have an upper limit on the X-ray luminosity \citep{callingham2021}. As we are calculating an upper limit on the mass-loss rate, we treat the upper limit on the X-ray luminosity as the highest value it could have.

We determine the mass-loss rate corresponding to each base density of the wind model. The highest mass-loss rate with an optical depth lower than one is taken to be the upper limit for this system. Our implementation of this method is available at \url{}
\begin{figure}
    \centering
    \includegraphics[width=0.95\columnwidth]{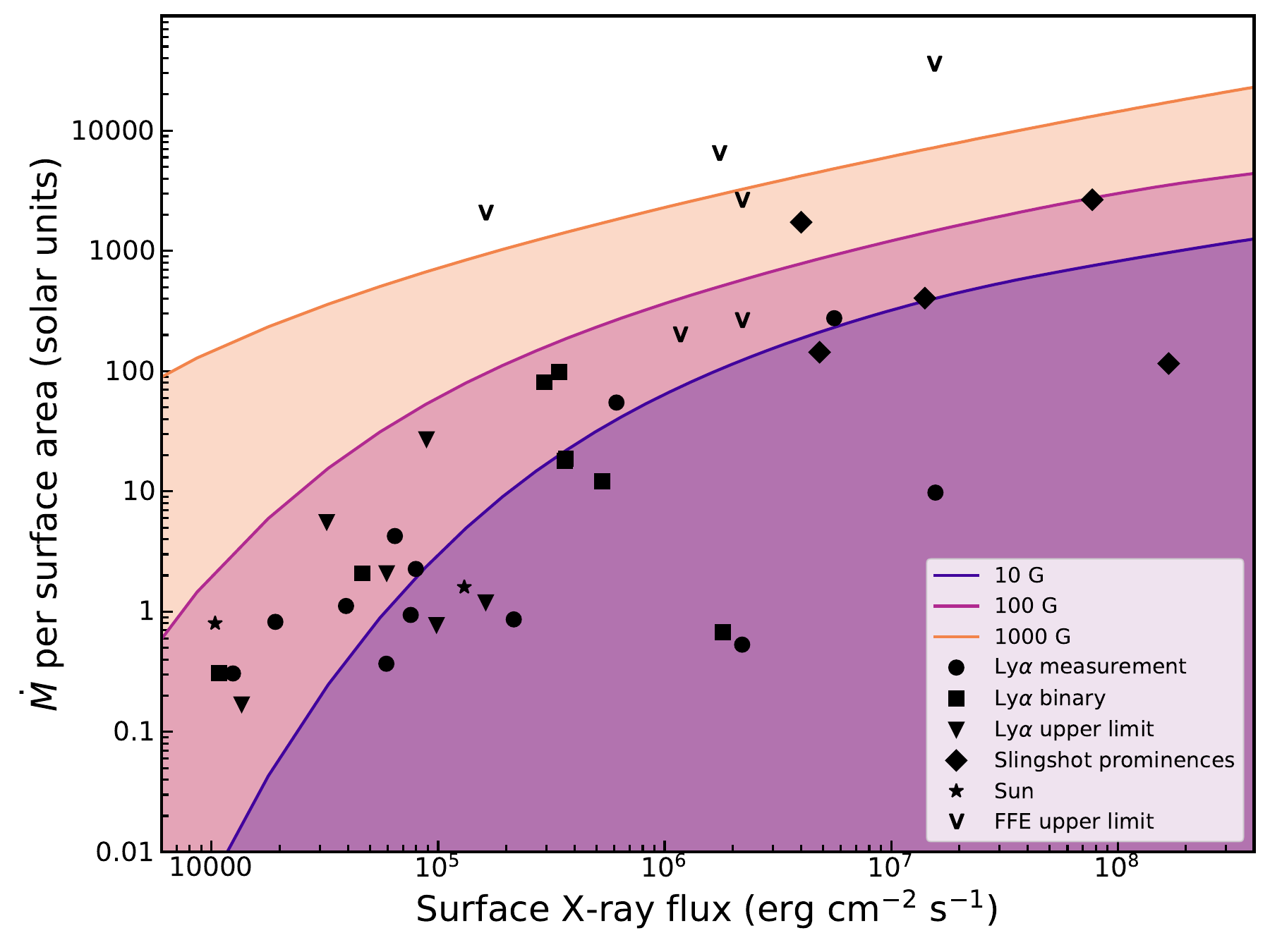}
\caption{The upper limit on the mass flux we find for a hypothetical star as a function of the surface X-ray flux, a proxy for coronal temperature, assuming ECMI emission was detected at 120\,MHz. The mass flux is in units of the solar mass flux, which we take to be $2\times10^{-14}$\,M$_{\odot}$\,yr$^{-1}$/(4$\pi$ R$_{\odot}^{2})$. We calculate the upper limits for three different dipole magnetic field strengths, indicated by the different colours. The shaded regions indicate the allowed regions for each corresponding dipole strength. The data points in black show previous measurements of mass-loss rates on low-mass stars from \citet{1996ApJ...462L..91L}, \citet{1996ApJ...460..976L}, \citet{gaidos}, \citet{Wood_2001}, \citet{wood2002}, \citet{Wood_2005}, \citet{Wood_2010}, \citet{Wood_2014}, \citet{Wood_2018}, \citet{2019MNRAS.482.2853J}, and \citet{Finley_2019} for reference. No error bars are plotted for the literature values as the uncertainties are not reported in the corresponding papers.}
\label{fig:test_star}
\end{figure}

To compare the sensitivity of our method to existing methods in literature, in Figure\,\ref{fig:test_star}, we show the upper limits on the mass flux (calculated as the mass-loss rate divided by the surface area of the star) we could obtain for a star with a mass of one solar mass and a radius of one solar radius, as a function of the surface X-ray flux (which determines the wind temperature in our model), assuming ECMI has been detected at 120\,MHz. The Low-Frequency Array \citep[LOFAR, ][]{vanHaarlem2013} commonly observes at 120\,MHz, the lowest frequency of its Two-metre all-sky survey, \citep[LoTSS; ][]{2019A&A...622A...1S}. 
We calculate the upper limits for three different dipole magnetic field strengths, shown in different colours. A stronger magnetic field allows the source to be farther away from the surface, resulting in a higher upper limit on the mass-loss rate. A weaker magnetic field, as shown by the 10\,G line, forces the source location to be very close to or even on the surface of the star, resulting in much stronger constraints. 

The dependence on surface X-ray flux is caused by the change in wind temperature. The temperature of the wind influences both the wind structure in the Parker model and the free-free opacity of the wind. For the same mass-loss rate, a higher temperature leads to a larger density scale height (i.e. a `puffier' corona) and a higher terminal wind velocity, so the same base density corresponds to a higher mass-loss rate. Combined with the decrease in opacity, increasing the temperature results in higher upper limits on the mass-loss rate.  Specifically, assuming a dipole field strength of 100\,G, a wind temperature of 1\,MK corresponds to a mass-loss rate of 6\,\mdotsun, while a coronal temperature of 10\,MK corresponds to a mass-loss rate of 3120\,\mdotsun.

Comparing the potential of our method to previous measurements of mass-loss rates of isolated low-mass main-sequence stars (Fig.\,\ref{fig:test_star}), we see that the upper limits our method can reach are in the same regime as the other measurements of stellar winds of low-mass stars. Even with the highest dipole magnetic field strength of 1\,kG, the upper limits are in the same regime as upper limits on the mass-loss rate in the literature.

\section{Application of the FFA method to LOFAR-detected M dwarfs}
\noindent Low-frequency detections of stars are ideal targets for our method, as they result in stronger and more physically informative constraints compared to higher-frequency radio detections. We therefore applied our methods to the sample of LOFAR-detected M\,dwarfs identified by \citet{callingham2021}. The sample consists of 19 M\,dwarfs detected at 120\,MHz. Even if the emitter is assumed to be the entire stellar disk, these detections have brightness temperatures in excess of $\sim 10^{14}$\,K, based on which \citet{callingham2021} identify the emission mechanism as ECMI. 

Out of the 19 stars in the sample, only three have magnetic field maps from ZDI, namely AD\,Leo \citep{2023A&A...676A..56B}, WX\,UMa \citep{2010MNRAS.407.2269M}, and GJ\,1151 \citep{lehmann}.
The magnetic fields of the other stars have not been measured, so we estimate an upper limit on the dipole component from their mass, using an empirical fit as described in Appendix\,\ref{app:mag_field}. The stars and their properties, including measured and estimated magnetic field strengths, are listed in Table\,\ref{tab:results}. For every star in our sample, we estimate an upper limit on the mass-loss rate with both a high and mean estimate for the dipole strength, also listed in Table\,\ref{tab:results}.

Three of the systems in our sample are known close binaries, where the components are expected to be roughly the same mass and spectral type. For these systems, we assume the radio emission is produced on the primary component. We calculate the coronal temperature using the total X-ray luminosity of the system as an upper limit. The resulting mass-loss rate is determined with a model of a single wind, and can be treated as an upper limit on the mass-loss rate of the primary star that could be overestimated due to a contribution of the second star.

We place the tightest limits on the mass-loss rates of low-mass stars from radio detections, easily reaching upper limits within a few orders of magnitude from the solar mass-loss rate when using high magnetic field estimates, and upper limits within one order of magnitude from the solar mass-loss rate using the mean magnetic field estimates.

\begin{table*}
\def\arraystretch{1.25}
\centering 
\caption{Properties of the sample of stars used in this work.}
    \resizebox{\linewidth}{!}{%
    \begin{tabular}{|l|ld{1.3}d{1.3}d{2.2}d{2.2}crrrr|}
    \hline \hline
         Name & Sp. type & \mc{M$_*$} & \mc{R$_*$ }& \mc{d} & \mc{$L_X$}& \mc{P$_{\mathrm{rot}}$}& \mc{$B_{0,\mathrm{low}}$} & \mc{\mdot$_{\mathrm{lim, low}}$ } & \mc{$B_{0,\mathrm{high}}$} & {\mdot$_{\mathrm{lim, high}}$ } \\
          & &  \mc{(M$_\odot$)} & \mc{(R$_\odot$) }& \mc{(pc)}&\mc{($10^{28}$erg s$^{-1}$) }& \mc{(d)} & \mc{(G)} & \mc{(\mdotsun) } & \mc{(G)} & {(\mdotsun) }  \\
         \hline 
DO Cep & M4.0 & 0.316 & 0.332 & 4.0 & 0.23 & 0.41 & 150 & {50} & 1500 & {350} \\
WX UMa$^*$ & M6.0 & 0.095 & 0.121 & 4.9 & 0.36 & 0.78 & - & - & 4300 & {260} \\
AD Leo$^*$ & M3.0 & 0.42 & 0.431 & 4.97 & 3.2 & 2.23 & - & - & 1000 & {990} \\
GJ 625 & M1.5 & 0.317 & 0.332 & 6.47 & 0.04 & $79.8\pm0.1$ & 150 & {12} & 1500 & {140} \\
GJ 1151$^*$ & M4.5 & 0.167 & 0.19 & 8.04 & 0.02 & 175.8$^{+3.2}_{-3.4}$ & - & - & 150 & {9} \\
GJ 450 & M1.5 & 0.46 & 0.474 & 8.76 & 0.66 & $23\pm1$ & 110 & {80} & 1100 & {560} \\
LP 169-22 & M5.5 & 0.111 & 0.138 & 10.4 & <0.03 & - & 320 & {75} & 3200 & {410} \\
CW UMa & M3.5 & 0.306 & 0.322 & 13.36 & 5.37 & 7.77 & 150 & {230} & 1500 & {1260} \\
HAT 182-00605 & M4.0 & 0.442 & 0.422 & 17.87 & 3.4 & 2.21 & 110 & {180} & 1100 & {1050} \\
LP 212-62 & M5.0 & 0.161 & 0.183 & 18.2 & 0.38 & 60.75 & 240 & {75} & 2400 & {410} \\
DG CVn$^\dagger$ & M4.0 & 0.56 & 0.567 & 18.29 & 10.72 & 0.11 & 94 & {350} & 940 & {1900} \\
GJ 3861 & M2.5 & 0.419 & 0.43 & 18.47 & 3.36 & - & 120 & {200} & 1200 & {1120} \\
CR Dra$^\dagger$ & M1.5 & 0.823 & 0.827 & 20.26 & 36.65 & 1.98 & 70 & {630} & 700 & {3400} \\
GJ 3729$^\dagger$ & M3.5 & 0.472 & 0.481 & 23.57 & 7.54 & 13.59 & 110 & {290} & 1100 & {1590} \\
G 240-45 & M4.0 & 0.125 & 0.162 & 27.59 & 0.02 & - & 300 & {22} & 3000 & {140} \\
2MASS J09481615+5114518 & M4.5 & 0.122 & 0.161 & 36.17 & 0.28 & - & 300 & {76} & 3000 & {400} \\
LP 259-39 & M5.0 & 0.173 & 0.202 & 36.93 & 18.7 & - & 230 & {250} & 2300 & {1500} \\
2MASS J10534129+5253040 & M4.0 & 0.408 & 0.379 & 45.19 & 28.01 & - & 120 & {370} & 1200 & {2160} \\
2MASS J14333139+3417472 & M5.0 & 0.101 & 0.136 & 47.84 & 0.83 & - & 350 & 100 & 3500 & 540 \\
         \hline
            \hline
    \end{tabular}%
    }
    
    \tablefoot{For each star, we list the stellar mass {M$_*$}, the stellar radius {R$_*$ }, the distance {d}, the X-ray luminosity {$L_X$}, the rotation period {P$_{\mathrm{rot}}$}, the low estimate for the magnetic field strength {$B_{0,\mathrm{low}}$} and the corresponding mass-loss rate limit {\mdot$_{\mathrm{lim, low}}$ }, and the high estimate for the magnetic field {$B_{0,\mathrm{high}}$} and its corresponding mass-loss rate limit {\mdot$_{\mathrm{lim, high}}$}. The systems marked with $\dagger$ are known close binary systems, for which we list the spectral type of the combined system. The stars marked with a $*$ have magnetic field measurements from ZDI. The stellar properties are taken from \citet{callingham2021}. The rotation period of GJ\,1151 is taken from \citet{lehmann}.}
    \label{tab:results}
\end{table*}

\begin{figure}
    \centering
    \includegraphics[width=0.95\columnwidth]{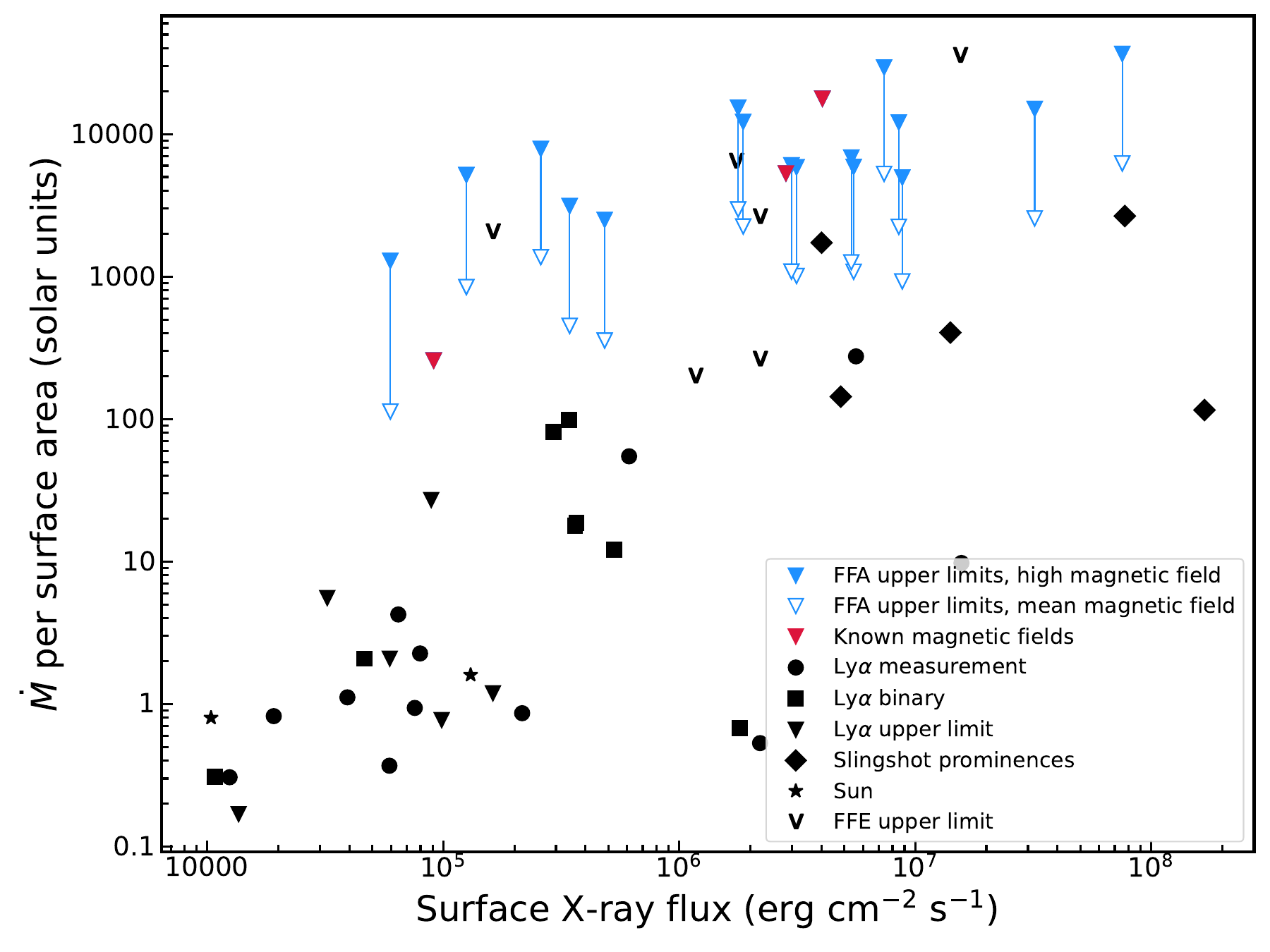}
    \caption{Mass flux as a function of X-ray surface flux. The filled and unfilled blue triangles show the upper limits on the mass-loss rate per unit surface area we find in this work for the stars with no magnetic field measurements, using respectively high and mean estimates for the magnetic field. The filled red triangles represent the stars for which the large-scale magnetic field is known. The mass flux is in units of the solar mass flux, which we take to be $2\times10^{-14}$\,M$_{\odot}$\,yr$^{-1}$/(4$\pi$ R$_{\odot}^{2})$. The black symbols show mass-loss rate measurements and estimates from \citet{1996ApJ...462L..91L}, \citet{1996ApJ...460..976L}, \citet{gaidos}, \citet{Wood_2001}, \citet{wood2002}, \citet{Wood_2005}, \citet{Wood_2010}, \citet{Wood_2014}, \citet{Wood_2018}, \citet{2019MNRAS.482.2853J}, and \citet{Finley_2019}. No error bars are plotted for the stellar literature values as the uncertainties are not reported in the corresponding papers.}
    \label{fig:all}
\end{figure}
We compare our mass-loss rate upper limits with the previously published state-of-the-art mass-loss rates from astrospheric absorption, slingshot prominences, and free-free emission (FFE). Most mass-loss rate measurements have been found with the astrospheric absorption method. In this method, the stellar wind is detected through an excess of Lyman alpha emission, blueshifted when compared to the ISM. The depth of absorption is linked to a mass-loss rate through hydrodynamical modelling. These models have been tested on the heliosphere and manage to reproduce it well \citep[e.g.][]{wood2002}. When applied to other stars, the velocity vector of the ISM and sometimes the temperature of the ISM are changed to match the conditions around the star, limiting it to stars within the Local Interstellar Cloud, where the ISM properties are well understood. The method is therefore only considered reliable out to a distance of $\sim$7\,pc \citep{wood2002}. Unfortunately, we cannot directly compare our upper limits to measurements through other methods, as there are no stars in our sample with mass-loss rate measurements or constraints through any other method. In the future, a direct comparison with astrospheric measurements would be particularly useful, as the different methods have different systematic errors.

In Figure\,\ref{fig:all}, we plot our upper limits on the mass flux of our stars, using both an upper limit on the magnetic field strength (referred to as the high magnetic field estimate) and an average value (referred to as the mean magnetic field estimate). We also plot currently known mass flux measurements as a function of their X-ray surface flux, only including single stars or binary systems that have been resolved individually in X-ray measurements, therefore excluding close binaries. Our limits are in the same regime as the previous measurements and upper limits, although a direct comparison is complicated by the lack of exact uncertainty estimates on the astrospheric absorption results. \citet{wood2004} state that the systematic uncertainties are difficult to quantify, but they estimate the uncertainty caused by the assumptions on the wind velocity and other factors to be roughly a factor of two.

We note that our method has an explicit dependence on the X-ray flux, as we use it to determine the temperature of the corona. The trend of increasing mass-loss rate with increasing X-ray flux is therefore not a physical effect. However, the upper limits are starting to probe a regime of high X-ray surface fluxes, where mass-loss rate measurements from traditional methods are sparse. \citet{Wood_2005} find that there might be a saturation in the increasing trend of mass-loss rates with surface X-ray flux, which they attribute to strong magnetic fields suppressing the stellar wind. The measurements based on stellar prominences by \citet{2019MNRAS.482.2853J} seem to contradict this, showing high mass-loss rates at high surface X-ray fluxes. With more radio detections of stars at low frequencies, our method can help fill this region of the parameter space and clarify the relationship between surface X-ray flux and the mass-loss rate.

\begin{figure}
    \centering
    \includegraphics[width=0.95\columnwidth]{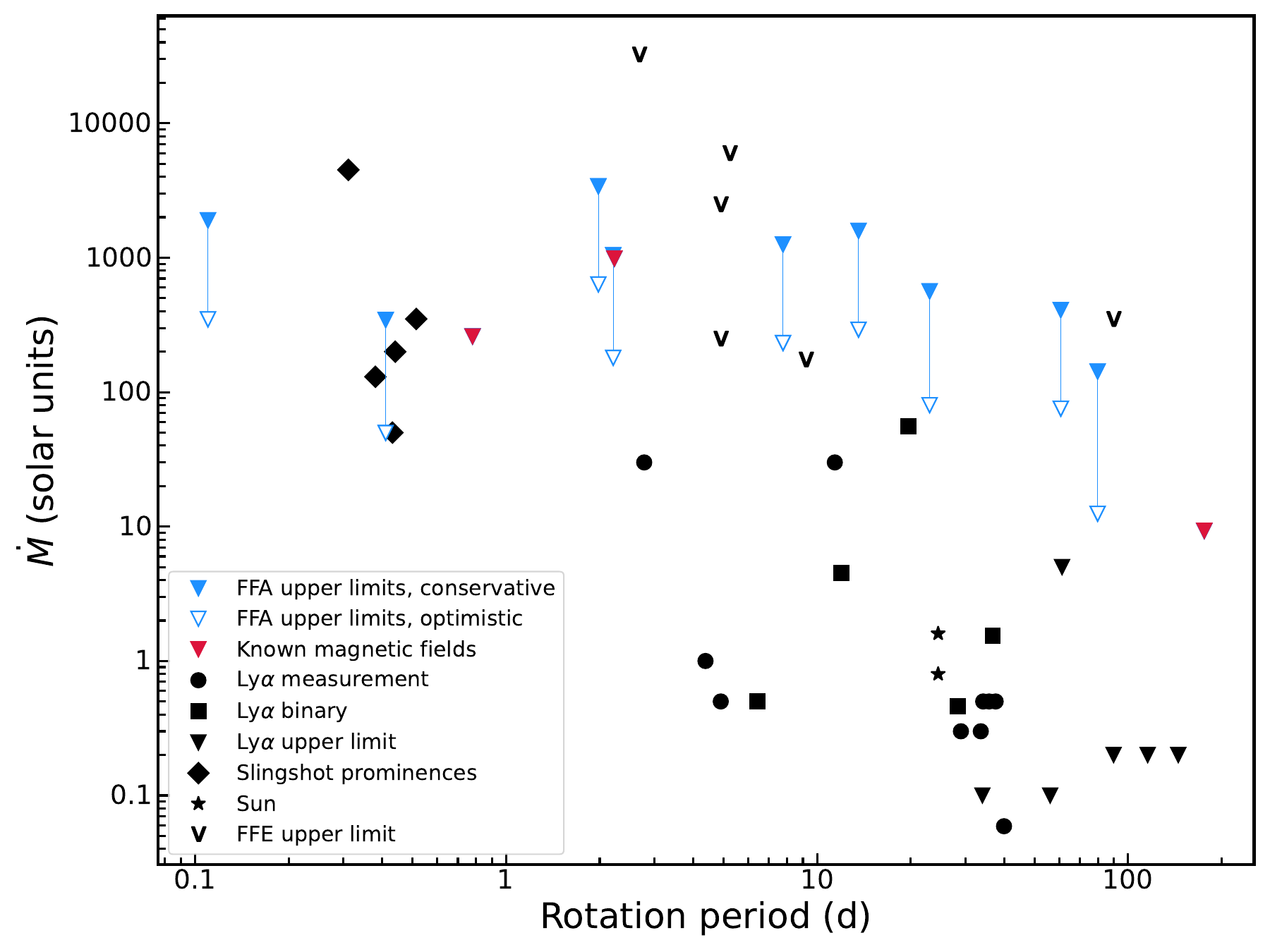}
\caption{The mass-loss rate as a function of rotation period. The filled and unfilled blue triangles show the upper limits on the mass-loss rate we find in this work for the stars with no magnetic field measurements, using respectively high and mean estimates for the magnetic field. The filled red triangles represent the stars for which the large-scale magnetic field is known.
    The black symbols show the same mass-loss rate measurements and estimates as Fig,\,\ref{fig:test_star} from \citet{1996ApJ...462L..91L}, \citet{1996ApJ...460..976L}, \citet{gaidos}, \citet{Wood_2001}, \citet{wood2002}, \citet{Wood_2005}, \citet{Wood_2010}, \citet{Wood_2014}, \citet{Wood_2018}, \citet{2019MNRAS.482.2853J}, and \citet{Finley_2019}. No error bars are plotted for the stellar literature values as the uncertainties are not reported in the corresponding papers.}
    \label{fig:Lx}
\end{figure}

We also plot our upper limits as a function of the rotation period, when known, in Figure\,\ref{fig:Lx}. Our sample covers a large range of rotation periods, ranging from 0.11 days (DG\,CVn) to 175.8$^{+3.2}_{-3.4}$ days (GJ\,1151).
We see a large scatter, with a slight trend to lower mass-loss rate upper limits with longer rotation periods. This could be an indirect effect of the X-ray flux in our model, as the rotation period and the X-ray luminosity are correlated \citep[e.g.][]{2011ApJ...743...48W}. A few of our upper limits are within the range of the previously measured mass-loss rates, implying that these stars could lie below the previously observed inverse correlation between the mass-loss rate and the rotation period \citep[e.g.][]{2019MNRAS.482.2853J}. 

We detect no stars with upper limits on their mass-loss rates less than a few times the solar mass-loss rate. Considering our conservative assumptions, these stars could be missing because their radio emission is absorbed. However, no detection of radio emission from a stellar system does not automatically mean that the radio emission is free-free absorbed. Instead, the star could be inactive at radio frequencies, the emission could be beamed away from our line of sight \citep[e.g.][]{kavanagh2023}, or another absorption mechanism may prevent radiation escape. This is the main drawback of our method: the lack of a radio detection does not lead to meaningful constraints.

Several of the stars in our sample have interesting properties, which we highlight in more detail below.

\subsection{GJ 1151}
\noindent GJ\,1151 has the weakest measured large-scale magnetic field in our sample \citep{lehmann}, leading to a stringent mass-loss rate limit. We use the highest observed magnetic field strength on the surface across all ZDI epochs, 150\,G, and assume that to be the dipole strength, to find a mass-loss rate upper limit of 9.3\,\mdotsun. If we instead use the lowest observed large-scale field strength, at 60\,G, we find an upper limit of 2.5\,\mdotsun, showing that we can reach upper limits on the order of the solar mass-loss rate with detailed magnetic field measurements.

\subsection{WX UMa}
\noindent WX\,UMa, another star with a mapped magnetic field \citep{2010MNRAS.407.2269M}, has been studied at radio frequencies in detail by \citet{davis-2019}, who find that the emission is produced in the ordinary magnetoionic mode. This implies that the emission is produced at the fundamental cyclotron frequency \citep[e.g.][]{dulk1985}. Given its known large-scale magnetic field strength of 4.3\,kG and the emission at the fundamental of the cyclotron frequency, we find an upper limit on the mass-loss rate of 260\,\mdotsun. 

WX\,UMa has a spectral type of M6, which, combined with its distance of 4.9\,pc, makes it difficult to determine the mass-loss rate through other means, as it is faint at optical wavelengths. Astrospheric absorption is limited to stars that are bright enough to detect the Ly$\alpha$ absorption signature. The latest spectral type for which this was attempted was Proxima\,Centauri, an M5.5 star, which was possible because of its proximity at 1.3\,pc \citep{wood_prox, gaia}, yet only resulted in an upper limit. WX\,UMa is therefore the latest spectral type with constraints on its mass-loss rate beyond our nearest neighbours.

\subsection{GJ 625}
\noindent Finally, we consider GJ\,625, the only system in our sample with a known planet in a close orbit. GJ\,625\,b is a super-Earth on the inner edge of the habitable zone, with an orbit of 14.6\,days \citep{2017A&A...605A..92S}. With estimates of both the mass-loss rate and the magnetic field structure of the star, we can determine the dynamic and magnetic pressure on the atmosphere of the planet, and therefore the probability of the atmosphere being stripped by the stellar wind.

The magnetic field of GJ\,625 has not been studied, so we use a range of values, from 10\,G to 4\,kG, to evaluate the fate of GJ\,625\,b's atmosphere. For each dipole strength, we calculate the upper limit on the mass-loss rate and use the corresponding density and velocity profile to calculate an upper limit on the ram pressure and magnetic pressure on the planet.
A limit on the ram and magnetic pressure on the atmosphere of a planet can be converted into a probability of retaining the atmosphere by assuming a model for how the wind and the atmosphere interact. Here, we use a simple model derived by \citet{2019A&A...630A..52R} (given by their Equation\,28), who base their estimate for the probability of retaining an atmosphere on the fraction of the planetary surface that is protected by the magnetic field. If we assume the magnetic field strength of GJ\,625\,b is at least 1\,G, we find that GJ\,625\,b has a 90\% probability or higher to retain its atmosphere if the stellar dipole strength is less than 30\,G.

This is the first time we have been able to infer the likelihood of retaining an atmosphere based on radio measurements of the system. This estimate is based on a simplistic recipe. To refine these estimates, we would require accurate measurements of the magnetic field strength of the star to improve the limits on the ram pressure and the magnetic pressure, along with a detailed model for atmospheric stripping that takes into account the properties of the exoplanet itself as well as those of the stellar wind.

\section{Comparison with free-free emission constraints}
\noindent To demonstrate how our method is an improvement over previous methods based on radio detections of free-free emission at higher frequencies, we first calculate the upper limit that could be found using free-free emission with the Jansky Very Large Array (JVLA). Assuming a sensitivity of 10\,$\mu$Jy at 45\,GHz, which can be achieved with a 4-hour observation, and a 5$\sigma$ detection threshold, we calculate the upper limits for one of our stars, GJ\,3861, following the method described in Appendix\,\ref{app:ffe}. The choice of frequency is based on maximising a potential FFE signal while minimising contamination from non-thermal processes, but is otherwise arbitrary. The best upper limit possibly found with the JVLA is 5000\,\mdotsun, whereas the upper limit we find is 1120\,\mdotsun using the high magnetic field estimate, and 200\,\mdotsun using the mean magnetic field estimate. The reason for this discrepancy is that our sensitivity to the mass-loss rate does not directly depend on distance. Only a detection is required. The free-free emission method, on the other hand, relies on the value of the flux density to determine the mass-loss rate, resulting in a lower sensitivity at higher distances.

\section{Conclusions}
\noindent Mass-loss rates of low-mass stars are notoriously difficult to measure due to their low densities, with the available methods relying on complex modelling of both the astrosphere and ISM. We present a new method to derive upper limits on the mass-loss rates of main-sequence stars, using detections of coherent radio emission to put an upper limit on the line-of-sight integrated density. We apply this method to 19 M\,dwarfs detected by LOFAR by \citet{callingham2021} and find competitive upper limits, with a sensitivity that does not depend on distance.

With future surveys of the low-frequency radio sky, the number of stars with upper limits derived with this method will grow significantly. In particular, once the LoTSS survey finishes, we can expect to detect around 100\,$\pm$\,30 M\,dwarfs \citep{callingham2021}.
Low-frequency and high sensitivity surveys in the Southern Hemisphere with the Square Kilometre Array\footnote{\href{https://www.skao.int/}{https://www.skao.int/}} in the future will at least double the number of M dwarfs detected at low frequencies. \citet{vlotss} estimate an expected yield of around 1000 detections of stellar coherent emission in a wide-field, 8-hour pointing, all-sky survey with SKA-Low. Since our method is independent of distance and spectral type, we expect to be able to apply it to a variety of spectral types and activity levels, providing a broad sample of mass-loss rate limits.

\section*{Data availability}
The models and code used in this work are available at \url{https://github.com/SBloot/stellar-winds-ECMI}.

\begin{acknowledgements}
\noindent We would like to thank J. F. Donati, O. Kochukhov, L. T. Lehmann, J. Morin, and V. See for providing the ZDI maps used to derive the estimates of magnetic fields. We also want to thank A. Vidotto for the useful discussions.

SB, HKV and RDK acknowledge funding from the Dutch Research Council (NWO) under the talent programme (Vidi grant VI.Vidi.203.093).  HKV acknowledges funding from the ERC starting grant `Stormchaser' (grant number 101042416).

This research made use of NASA's Astrophysics Data System, the \textsc{IPython} package \citep{PER-GRA:2007}; \textsc{SciPy} \citep{scipy}; \textsc{Matplotlib}, a \textsc{Python} library for publication-quality graphics \citep{Hunter:2007}; \textsc{Astropy}, a community-developed core \textsc{Python} package for astronomy \citep{2013A&A...558A..33A}; and \textsc{NumPy} \citep{van2011numpy}. 
\end{acknowledgements}

\bibliographystyle{aasjournal}
\bibliography{stellarwinds} 

\newpage
\appendix

\section{Estimating the magnetic field of M dwarfs}
\label{app:mag_field}
\begin{figure}[b]
    \centering
    \includegraphics[width=0.95\columnwidth]{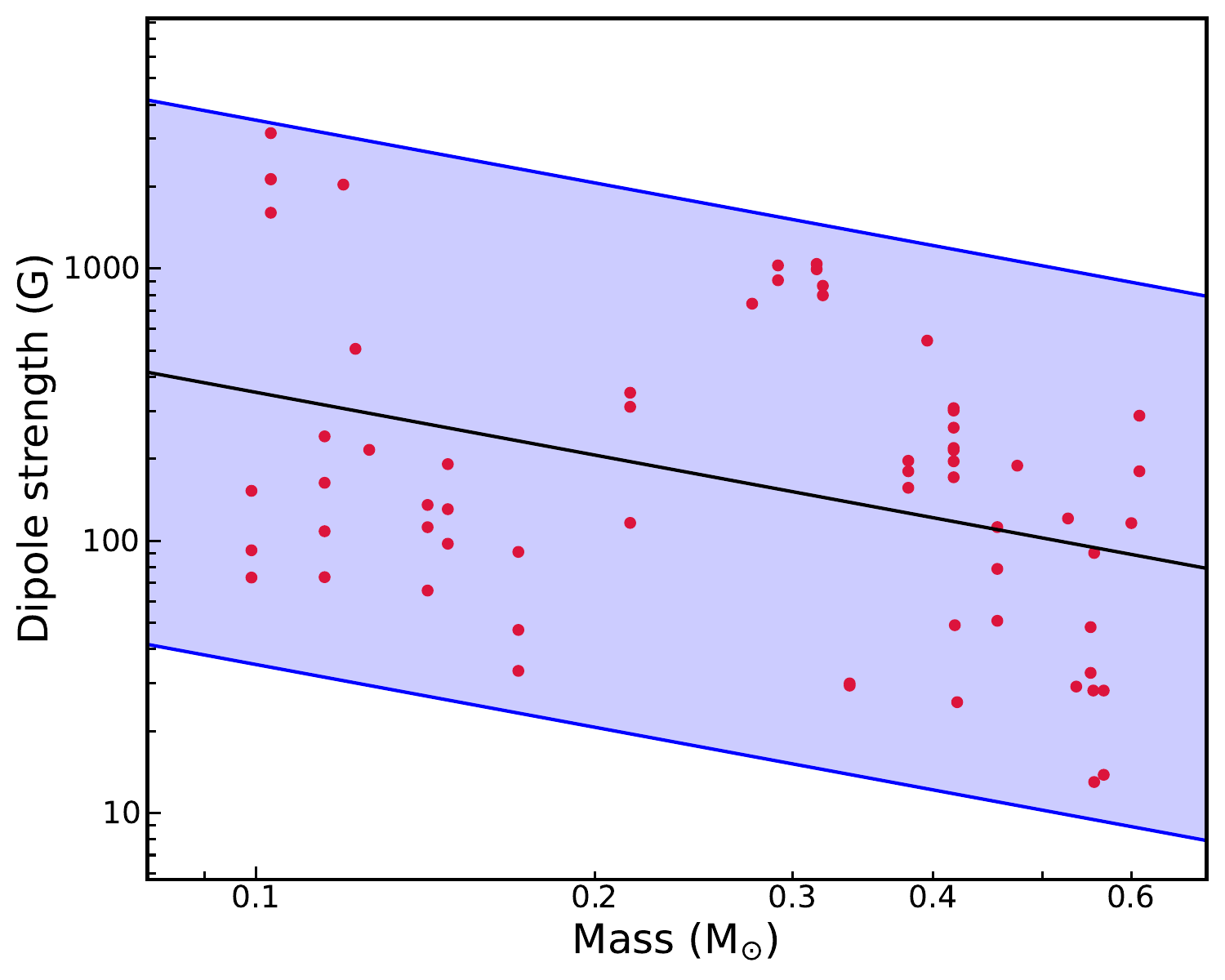}
    \caption{The dipole field strength as a function of mass, for a sample of M dwarfs with ZDI maps, shown as red points. The black line shows the best fit to the data, which we treat as the mean expected value. The blue lines show the upper and lower end of the possible range of dipole strengths, based on the scatter in the sample. The shaded region shows the full range of expected dipole strengths as a function of mass.}
    \label{fig:mass}
\end{figure}

For most radio-detected stars, we do not have detailed ZDI maps to show the strength and structure of the magnetic field. We therefore require a method to estimate the possible range of dipole magnetic field strengths for a star from known parameters. We collected a sample of M dwarfs with ZDI maps from \citet{2006Sci...311..633D}, \citet{2008MNRAS.390..545D}, \citet{2008MNRAS.390..567M}, \citet{2008MNRAS.384...77M}, \citet{2010MNRAS.407.2269M}, \citet{2016MNRAS.461.1465H}, \citet{2017ApJ...835L...4K}, \citet{2017MNRAS.472.4563M}, \citet{2019ApJ...873...69K}, \citet{2021MNRAS.500.1844K}, \citet{2023A&A...676A.139B}, \citet{2023A&A...678A.207M}, \citet{2023A&A...676A..56B}, and \citet{lehmann}, and looked for correlations between the dipole strengths and other properties. We did not include Zeeman broadening measurements, as they may represent the strongest small-scale structures on the surface instead of the large-scale dipole. The strongest correlation is between the dipole magnetic field strength and the stellar mass, as shown in Figure\,\ref{fig:mass}. To describe the range of values covered by this sample, we fit a power law to the data in log space. The best fit is given by \begin{equation}
   \log_{10}(B_{\mathrm{dip}})= -0.765\log_{10}(M_*)+1.78,
\end{equation}
where $B_{\mathrm{dip}}$ is the dipole field strength in Gauss, and $M_*$ is the stellar mass in solar masses.
We treat this fit as the mean expected value for the dipole magnetic field strength at a given mass. We determine the range of possible values at a given mass by taking a $2\sigma$ offset (in log space) from this mean expected value. The scatter in the sample in log space after subtracting the mean is 0.5, so we add a $\pm1$ offset to the best fit to define the range of possible values. For our conservative (high) estimates of the mass-loss rate, we use the upper limit of this range of potential values. This may overestimate the dipole magnetic field strength and therefore the final upper limit on the mass-loss rate, but we cannot determine this a priori. More ZDI measurements of stars of radio interest are required to determine the actual magnetic field strength. For the optimistic (mean) estimates, we use the mean expected value. 

\section{Free-free emission}
\label{app:ffe}
\begin{figure}[h]
    \centering
    \includegraphics[width=0.8\linewidth]{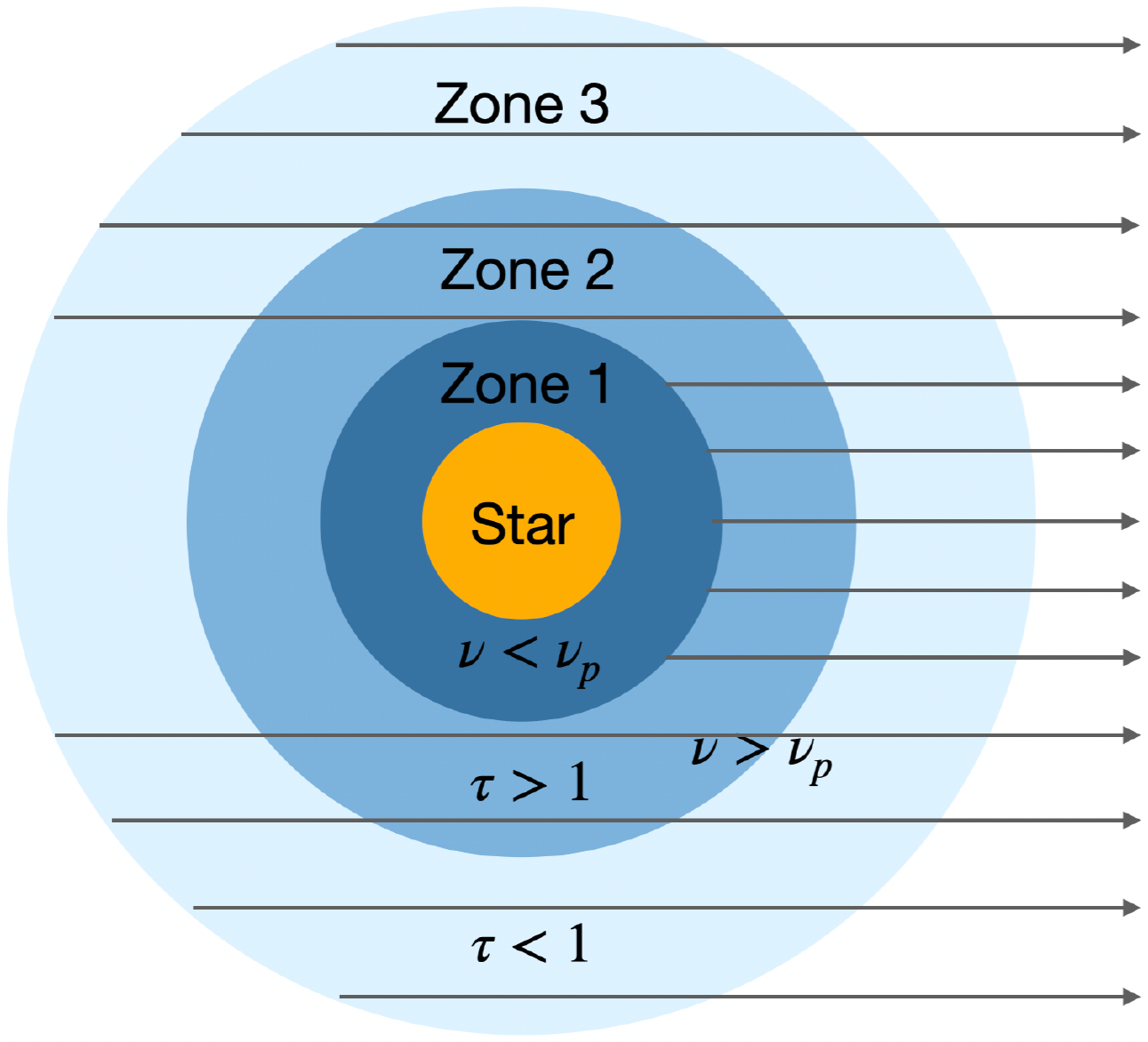}
    \caption{Geometry for the free-free emission calculation code showing the different zones and the path of the rays on which the radiative transfer equation is numerically integrated. The stellar meridian and the observer lie in the plane of the page.}
    \label{fig:xx}
\end{figure}

To compute the free-free emission flux of a stellar wind, we assume a spherically symmetric Parker wind for a given stellar mass, radius, mass-loss rate, and coronal temperature, assuming an isothermal wind. We divide the corona into three shells: zone\,1, the innermost shell where the observation frequency is below the plasma frequency from where no radiation can escape, zone\,2, the intermediate shell whose free-free optical depth is $\tau\gg 1$ which means the radiation brightness temperature at the outer edge of the shell is always equal to the kinetic temperature of the corona, and zone\,3, where the approximation that the medium is optically thick cannot be made and we numerically integrate the equation of radiative transfer. Due to the spherical symmetry of the problem, we only trace rays passing through the stellar meridian (Figure\,\ref{fig:xx}) to speed up computations. 
\end{document}